\documentclass[twocolumn,preprintnumbers,amsmath,amssymb]{revtex4}
\usepackage{epsfig}
\usepackage{color}
\usepackage{dsfont}
\usepackage[colorlinks,urlcolor=blue,linkcolor=blue,citecolor=red]{hyperref}
\begin{document}
\setlength{\voffset}{1.0cm}
\title{Chiral spiral in the presence of chiral imbalance}
\author{Michael Thies\footnote{michael.thies@gravity.fau.de}}
\affiliation{Institut f\"ur  Theoretische Physik, Universit\"at Erlangen-N\"urnberg, D-91058, Erlangen, Germany}
\date{\today}

\begin{abstract}
The phase diagram of the two-dimensional Nambu--Jona-Lasinio (or chiral Gross-Neveu) model is characterized by an order parameter in the form
of a chiral spiral. Its radius vanishes at a critical temperature, its period depends only on the chemical potential. We generalize these findings
to chirally imbalanced systems by including a chiral chemical potential $\mu_5$. The relationship between the present, static approach and 
a previous, time dependent one is traced back to a half-local symmetry which the NJL$_2$ model shares with massless Dirac fermions, but which has been 
neglected so far. The structure of chiral spiral matter is further elucidated by computing fermion and antifermion momentum distribution functions,
using a Bogoliubov transformation.
\end{abstract}

\maketitle

\section{Introduction}
\label{sect1}

Much can be learned from exactly solvable quantum field theories, for instance the Gross-Neveu model \cite{1}. Here we reconsider the version with
U(1)$\times$U(1) chiral symmetry, i.e., the Nambu--Jona-Lasinio model \cite{2} in 1+1 dimensions (NJL$_2$) with Lagrangian
\begin{equation}
{\cal L}_{{\rm NJL}_2} = \bar{\psi} i \partial \!\!\!/ \psi + \frac{g^2}{2}\left[ (\bar{\psi}\psi)^2 + ( \bar{\psi} i \gamma_5 \psi )^2 \right].
\label{I1}
\end{equation}
The model is endowed with a U($N$) flavor symmetry, thus generating a useful expansion parameter, $1/N$. We suppress flavor indices in (\ref{I1})
as usual ($\bar{\psi}\psi = \sum_{i=1}^{N} \bar{\psi}_i \psi_i$ etc.). Among many interesting results, the most striking one is perhaps the structure of hot and
dense matter, resulting in a remarkably simple phase diagram as a function of temperature $T$ and chemical potential $\mu$ \cite{3,4}.
At zero temperature, applying a linearly $x$-dependent, local chiral rotation to the vacuum spinors shifts the spectrum rigidly up, thereby
pulling occupied fermion levels out from the bottom of the Dirac sea --- a manifestation of the chiral anomaly in 1+1 dimensions. At the same time, the order
parameter changes from a constant mass to a chiral spiral with helical symmetry \cite{5}. As an ultraviolet (UV) effect, the axial anomaly gives rise to temperature
independent changes of thermodynamic observables upon changing $\mu$. Surprisingly, several global observables mimick a free Fermi gas of massless particles,
although the physical fermions do acquire a ($T$-dependent) dynamical mass due to spontaneous breaking of chiral symmetry, at least below a certain critical temperature.

During the last few years, there has been increased interest in chirally imbalanced matter with unequal densities of left and right handed quarks.
The motivation stems primarily from the chiral magnetic effect and other potentially observable signals from the chiral anomaly in ultra-relativistic heavy ion collisions,
compact stars or quasi-relativistic condensed matter systems (for a recent review, see \cite{6}). This incites us to revisit the chiral spiral in the context of chiral
imbalance, even if this is not directly relevant for real physical systems.

In 1+1 dimensions, the axial charge density $\rho_5= \bar{\psi} \gamma^0 \gamma_5 \psi$ coincides with the vector current density $j=\bar{\psi} \gamma^1 \psi$.
Therefore a chirally imbalanced system carries a non-vanishing current density. Its magnitude can be controlled by introducing an axial chemical potential $\mu_5$,
conjugate to the axial charge $Q_5$.

We are aware of a variational calculation of the NJL model with isospin [pseudoscalar interaction term in (\ref{I1}) replaced by
$(\bar{\psi} i \gamma_5 \vec{\tau} \psi)^2$] in 1+1 dimensions, including a chiral chemical potential \cite{7,8}. 
The two references most closely related to the present work are Refs.~\cite{9} and \cite{10}.
Ref.~\cite{9} considers a more complicated four-fermion model, where the Lagrangian (\ref{I1}) is augmented by  a term inducing Cooper pairing. 
The full phase diagram in ($T,\mu,\mu_5$)-space is determined in a variational calculation, using as ansatz potentials of chiral spiral type. We shall compare this 
approach to the present one at the end of Sect.~\ref{sect2}.
As far as the simpler model (\ref{I1}) in 1+1 dimensions is concerned, there is some recent work on the imbalanced system at $T=0$ \cite{10}. Since a system
with finite fermion density acquires a current density if viewed by an observer in a moving inertial frame,
it was argued that the mean fields of systems with $\mu_5=0$ and $\mu_5 \neq 0$ should be related by a Lorentz transformation. Boosting the chiral spiral
necessarily yields space and time dependent condensates. There is nothing wrong with this in principle, but it makes the transition to finite
temperature thermodynamics more difficult. 

Here we propose another, more straightforward approach to the NJL$_2$ model at finite $T,\mu,\mu_5$, and present a full, static Hartree-Fock (HF) solution. As a by product,
we also clarify the physics content of the chiral spiral state of matter by evaluating new observables, namely momentum distributions of ``quarks" and ``antiquarks".
The relation between the present static approach and the time dependent one of Ref. \cite{10} will also be addressed.

This paper is organized as follows: In Sect.~\ref{sect2}, we generalize the chiral spiral solution to $\mu_5 \neq 0$ and derive the full NJL$_2$ phase diagram.
In Sec.~\ref{sect3}, we discuss the difference between this approach and a previous, time dependent one. In Sect.~\ref{sect4} we give a more physical picture
of chiral spiral matter by computing quark and antiquark momentum distributions. We finish with a short summary and our conclusions in Sect.~\ref{sect5}. 

\section{Full phase diagram with chiral chemical potential}
\label{sect2}

The purpose of the present section is to map out the phase diagram of the NJL$_2$ model (\ref{I1}) as a function of chemical potential $\mu$, chiral chemical potential
$\mu_5$ and temperature $T$. We restrict ourselves to the 't~Hooft limit ($N\to \infty, Ng^2 = $ const.) where the relativistic version of thermal Hartree-Fock (HF) theory
is believed to become exact. The calculation follows closely the one at $\mu_5=0$ in \cite{3}, thus we shall move rather fast through some of the formal steps. We work with the 
grand canonical ensemble, generalized to two chemical potentials. The HF equations read
\begin{equation}
\left( - \gamma_5 i \partial_x - \mu - \mu_5 \gamma_5 + \gamma^0 S+ i \gamma^1 P\right) \psi_{\alpha} = \omega_{\alpha} \psi_{\alpha},
\label{A1}
\end{equation}
supplemented by the (finite temperature) self-consistency conditions
\begin{eqnarray}
S & = & -  g^2 \langle \bar{\psi}\psi \rangle = - Ng^2 \sum_{\alpha} \bar{\psi}_{\alpha} \psi_{\alpha} \frac{1}{e^{\beta \omega_{\alpha}}+1},
\label{A2} \\
P & = & - g^2 \langle \bar{\psi} i \gamma_5 \psi \rangle =  - Ng^2 \sum_{\alpha} \bar{\psi}_{\alpha} i \gamma_5 \psi_{\alpha}  \frac{1}{e^{\beta \omega_{\alpha}}+1}.
\nonumber
\end{eqnarray}
Next we transform away the chemical potentials from Eq.~(\ref{A1}) by the following local chiral transformation,
\begin{equation}
\psi_{\alpha} = e^{i \mu x \gamma_5} e^{i\mu_5 x} \phi_{\alpha}.
\label{A3}
\end{equation}
Due to the $\gamma_5$ matrix, this will also affect the scalar and pseudoscalar potentials $S,P$.
The Dirac equation satisfied by $\phi_{\alpha}$ becomes
\begin{equation}
\left( - \gamma_5 i \partial_x + \gamma^0 \tilde{S}+ i \gamma^1 \tilde{P}\right) \phi_{\alpha} = \omega_{\alpha} \phi_{\alpha}
\label{A4}
\end{equation}
with
\begin{equation}
\left( \begin{array}{c} \tilde{S} \\ \tilde{P} \end{array} \right) = \left( \begin{array}{rr} \cos 2\mu x & - \sin 2\mu x \\  \sin 2\mu x & \cos 2\mu x \end{array} \right) 
\left( \begin{array}{c} S \\ P \end{array} \right) .
\label{A5}
\end{equation}
As  $\langle \bar{\psi}\psi \rangle,   \langle \bar{\psi} i \gamma_5 \psi \rangle$ transform in the same way as $S,P$ under (\ref{A3}), 
the self-consistency relations (\ref{A2}) go over into
\begin{eqnarray}
\tilde{S} & = & -  g^2 \langle \bar{\phi}\phi \rangle = - Ng^2 \sum_{\alpha} \bar{\phi}_{\alpha} \phi_{\alpha} \frac{1}{e^{\beta \omega_{\alpha}}+1},
\label{A6} \\
\tilde{P} & = & - g^2 \langle \bar{\phi} i \gamma_5 \phi \rangle =  - Ng^2 \sum_{\alpha} \bar{\phi}_{\alpha} i \gamma_5 \phi_{\alpha}  \frac{1}{e^{\beta \omega_{\alpha}}+1}.
\nonumber
\end{eqnarray}
Eqs.~(\ref{A4},\ref{A6}) are just the thermal HF equations for the NJL$_2$ model at zero chemical potentials, where we already know the answer.
For the standard choice of the global chiral phase,
\begin{equation}
\tilde{S} = m, \quad \tilde{P} = 0,
\label{A7}
\end{equation} 
with $m$ the ($T$-dependent) dynamical fermion mass.
Thus the original HF problem has been reduced to the free, massive Dirac equation with well-known spinors and spectrum 
\begin{equation}
\omega_k = \pm \epsilon_k  = \pm \sqrt{k^2+m^2} .
\label{A7a}
\end{equation}
From (\ref{A5}) and (\ref{A7}) we then infer that the mean fields at finite chemical potentials $\mu, \mu_5$ are
\begin{eqnarray}
S  =  m \cos 2\mu x & ,&  \quad P =- m \sin 2 \mu x,
\nonumber \\
\Delta & = &  S-iP = m e^{2i\mu x}.
\label{A8}
\end{eqnarray}
This is identical to the standard result for the chiral spiral, independently of $\mu_5$. 

For the sake of completeness, let us briefly recall how to minimize the effective potential (free energy) density at zero chemical potentials. One starts from the standard expression 
\begin{equation}
\frac{{\cal V}_{\rm eff}}{N} = \frac{m^2}{2Ng^2} - \int_{-\Lambda}^{\Lambda} \frac{dk}{2\pi} \epsilon_k - \frac{2}{\beta} \int \frac{dk}{2\pi} \ln \left( 1+e^{-\beta \epsilon_k}\right)
\label{A9}
\end{equation}
still containing the bare coupling constant $Ng^2$ and an UV cutoff $\Lambda$. This has to be minimized with respect to $m$, the dynamical, $T$-dependent
fermion mass. At $T=0$ in particular, denoting the mass at the minimum by $m_0$, one finds the vacuum gap equation
\begin{equation}
\frac{\pi}{Ng^2} - \ln \frac{2\Lambda}{m_0} = 0.
\label{A10}
\end{equation}
It can be used to renormalize the effective potential, trading the dimensionless bare coupling against a dimensionful scale parameter $m_0$,
\begin{equation}
\frac{{\cal V}_{\rm eff}}{N} = \frac{m^2}{4\pi} \left( \ln \frac{m^2}{m_0^2}-1 \right) + \frac{m_0^2}{4\pi} -  \frac{2}{\beta} \int \frac{dk}{2\pi} \ln \left( 1+e^{-\beta \epsilon_k}\right).
\label{A11}
\end{equation}
This expression has been normalized to 0 at $T=0$ by adding the term $\sim m_0^2$.
If one then minimizes expression (\ref{A11}) with respect to $m$, one finds that $m$ decreases monotonically, vanishing at the critical
temperature $T_c= m_0 e^{\gamma_E}/\pi$ ($\gamma_E =$ Euler constant), cf. the original reference \cite{11}. 

\begin{figure}
\begin{center}
\epsfig{file=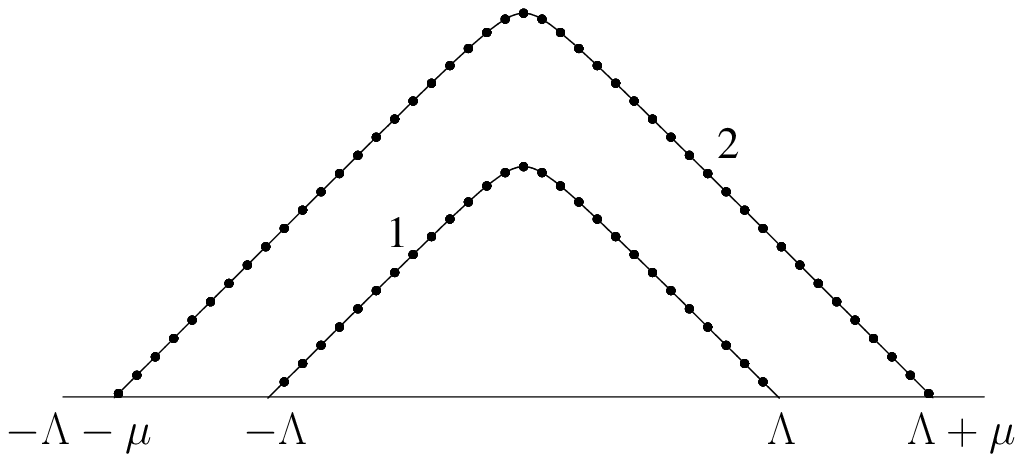,width=8cm,angle=0}
\caption{Spectrum and filling of single particle states versus $k$ in HF approximation for vacuum ( $- \epsilon_k=-\sqrt{m^2+k^2}$, curve 1) and chiral spiral ($-\epsilon_k+\mu$, curve 2).
The fermion density is increased through the chiral anomaly.}
\label{fig1}
\end{center}
\end{figure}

What changes at finite $\mu, \mu_5$? Owing to the transformation (\ref{A3}), we know that expression (\ref{A9}) remains valid for finite chemical potential, at least formally. 
Since the double counting correction [the first term in ${\cal V}_{\rm eff}$, Eq.~(\ref{A9})] and the fermion spectrum are independent of the chemical potentials, the only place
left where these could affect the effective potential is through the UV cutoff in the 2nd term. The 3rd term, the only one depending on temperature, is also independent of the chemical potentials.
So let us focus on the $T=0$ limit for the moment. We shall come back to the full phase diagram later on. In the chirally symmetric case ($\mu_5=0$), it was
argued that the UV cutoff has to be changed from $\Lambda$ to $\Lambda+\mu$ so as to keep the cutoff in single particle energies constant (see Fig.~\ref{fig1}). 
This affects the number of occupied levels, resulting in the fermion density 
\begin{equation}
\frac{\rho}{N} = \int_{-\Lambda-\mu}^{\Lambda+\mu} \frac{dk}{2\pi} = \frac{\Lambda}{\pi} + \frac{\mu}{\pi}.
\label{A12}
\end{equation}
The divergent density of the Dirac sea $\sim \Lambda$ should of course be subtracted. The conventional picture of a Fermi gas is 
a state where positive energy levels in the ``Fermi sphere" $|k|<k_f$ are filled in addition to the negative energy Dirac sea.
Here, instead, one works exclusively with negative energy levels, with the benefit that the gap always remains at the Fermi surface 
(``Peierls instability" in condensed matter systems \cite{12}).
With this scenario in mind, it is now almost trivial to treat chiral imbalance as well. In the UV region where extra fermions are added
(for matter, $\mu>0$) or removed (for antimatter, $\mu<0$), we are dealing with states of definite chirality, since the mass becomes irrelevant. The chirality of a 
negative energy state $\phi_k^{(-)}$ is 
\begin{equation}
\phi_k^{(-)\dagger} \gamma_5 \phi_k^{(-)} = - \frac{k}{\epsilon_k} \to - {\rm sgn}(k) \quad {\rm for} \ \ |k| \to \infty
\label{A13}
\end{equation}
as can be verified with the explicit spinors below, cf. Eqs.~(\ref{C6},\ref{C7}).
Thus in order to change the density of right handed fermions, we only need to replace the lower cutoff $-(\Lambda+\mu)$ by $-(\Lambda+\mu_R)$. 
To change the density of left handed fermions we replace the upper cutoff $\Lambda+\mu$ by $\Lambda+\mu_L$. Here, the chemical potentials 
\begin{equation}
\mu_R = \mu + \mu_5, \quad \mu_L = \mu - \mu_5
\label{A14}
\end{equation}
are chemical potentials conjugate to the densities of right- and left handed quarks.
At $T=0$, the difference between the effective potentials at $\mu, \mu_5$ and at $\mu=\mu_5=0$ at $T=0$ [cf. Eq.~(\ref{A11})] is therefore 
\begin{eqnarray}
 \lim_{\beta \to \infty}&  & \frac{{\cal V}_{\rm eff}(\beta, \mu, \mu_5) - {\cal V}_{\rm eff}(\beta, 0, 0)}{N}  
\nonumber \\
& = & - \int_{-\Lambda-\mu_R}^{\Lambda+\mu_L} \frac{dk}{2\pi} \epsilon_k +  \int_{-\Lambda}^{\Lambda} \frac{dk}{2\pi} \epsilon_k
\nonumber \\
& = & - \frac{\Lambda \mu}{\pi} - \frac{\mu^2}{2\pi} - \frac{\mu_5^2}{2\pi}.
\label{A15} 
\end{eqnarray}
The densities $\rho, \rho_5$ can either be obtained via thermodynamic identities 
\begin{eqnarray}
\frac{\rho}{N} & = & - \frac{\partial}{\partial \mu} \frac{{\cal V}_{\rm eff}(\beta, \mu, \mu_5)}{N} = \frac{\Lambda}{\pi} +  \frac{\mu}{\pi},
\nonumber \\
\frac{\rho_5}{N} & = & - \frac{\partial}{\partial \mu_5} \frac{{\cal V}_{\rm eff}(\beta, \mu, \mu_5)}{N} = \frac{\mu_5}{\pi},
\label{A16}
\end{eqnarray}
or else by direct computation
\begin{eqnarray}
\frac{\rho}{N} & = & \int_{-\Lambda-\mu_R}^{\Lambda+ \mu_L} \frac{dk}{2\pi} =  \frac{\Lambda}{\pi} +  \frac{\mu}{\pi},
\nonumber \\
\frac{\rho_5}{N} & = & \int_{-\Lambda-\mu_R}^{\Lambda+ \mu_L} \frac{dk}{2\pi} \left( - \frac{k}{\epsilon_k} \right)  =   \frac{\mu_5}{\pi},
\label{A17}
\end{eqnarray}
with consistent results. The change in energy density at $T=0$ in turn is given by
\begin{eqnarray}
\frac{{\cal E}(\mu,\mu_5) -{\cal E}(0,0)}{N} & = &  \lim_{\beta \to \infty} \frac{{\cal V}_{\rm eff}(\beta, \mu, \mu_5)  - {\cal V}_{\rm eff}(\beta, 0, 0)}{N}
\nonumber\\
& & + \mu \frac{\rho}{N} + \mu_5 \frac{\rho_5}{N}
\nonumber \\
& = & \frac{\mu^2}{2\pi} + \frac{\mu_5^2}{2\pi}.
\label{A18}
\end{eqnarray}
Alternatively, we can write ${\cal E}/N$ as follows
\begin{equation}
\frac{{\cal E}(\mu,\mu_5)}{N} = \int_{-\Lambda-\mu_R}^{\Lambda+ \mu_L} \frac{dk}{2\pi} \left( - \epsilon_k + \mu - \mu_5 \frac{k}{e_k} \right)
\label{A19}
\end{equation}
where the integrand is the expectation value
\begin{equation}
\phi_k^{(-) \dagger} \left( -i \gamma_5 \partial_x + \gamma^0 m + \mu + \mu_5 \gamma_5 \right) \phi_k^{(-)}.
\label{A20}
\end{equation}
\begin{figure}
\begin{center}
\epsfig{file=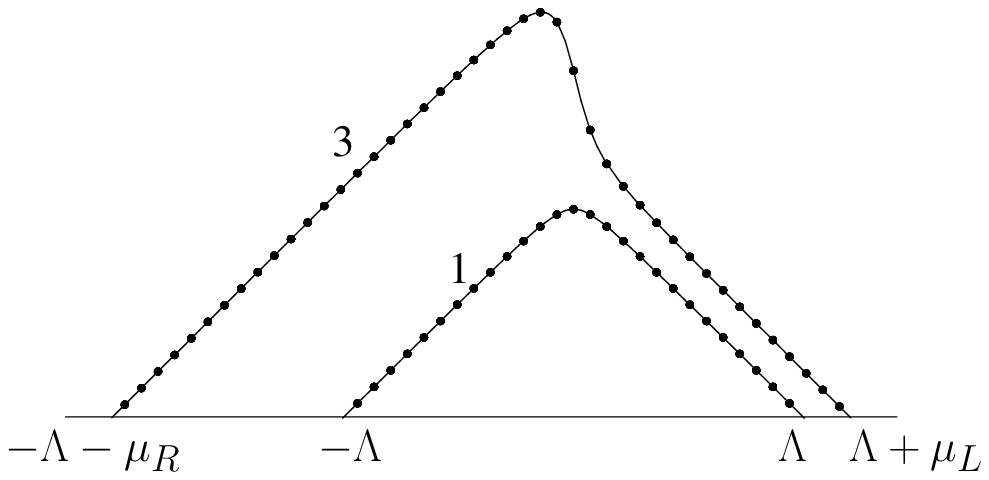,width=8cm,angle=0}
\caption{Like Fig.~\ref{fig1}, but for chirally imbalanced state. Curve 1: vacuum, curve 3: HF state with different chemical potentials $\mu_R, \mu_L$. The 
expectation value of the single particle Hamiltonian, Eq.~\ref{A20}, is plotted against $k$.}
\label{fig2}
\end{center}
\end{figure}
Since $\phi_k^{(-)}$ is an eigenvector of the grand canonical HF Hamiltonian but not of the HF Hamiltonian, it is the expectation value which enters here.
Eq.~(\ref{A19}) enables us to illustrate how the picture of Fig.~\ref{fig1} changes upon introducing $\mu_5$, see Fig.~\ref{fig2}. The cutoff in single particle
energies (or rather expectation values of the single particle Hamiltonian) is kept fixed while changing $\mu,\mu_5$. Whereas the chemical potential $\mu$ induces 
a rigid upward shift of the vacuum picture, $\mu_5$ gives rise to a sideways shift and a distortion.

\begin{figure}
\begin{center}
\epsfig{file=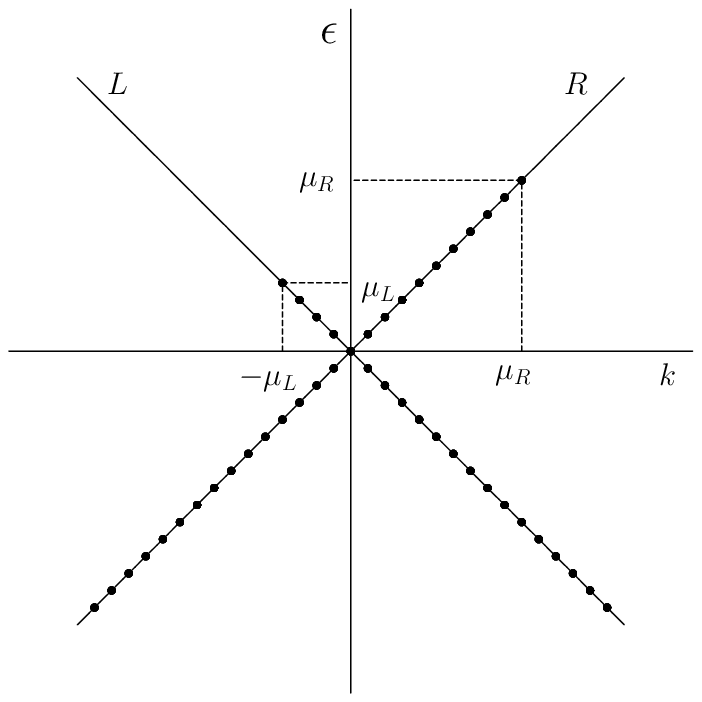,width=8cm,angle=0}
\caption{Chirally imbalanced state for free, massless Dirac fermions, illustrating the origin of the results shown in Eq.~(\ref{A24}).}
\label{fig3}
\end{center}
\end{figure}

A new observable at $\mu_5 \neq 0$ is the momentum density. Since the single particle states are not momentum eigenstates
(due to breaking of translational invariance), we cannot simply sum up eigenvalues but need again the expectation values
\begin{eqnarray}
\frac{\cal P}{N} & = & \sum_{\alpha}^{\rm occ} \psi_{\alpha}^{\dagger}\frac{1}{i} \partial_x \psi_{\alpha}
\label{A21} \\
& = &  \sum_{\alpha}^{\rm occ} \phi_{\alpha}^{\dagger}\frac{1}{i} \partial_x \phi_{\alpha} + \mu  \sum_{\alpha}^{\rm occ} \phi_{\alpha}^{\dagger}\gamma_5 \phi_{\alpha}
+  \mu_5  \sum_{\alpha}^{\rm occ} \phi_{\alpha}^{\dagger}\phi_{\alpha}.
\nonumber
\end{eqnarray}
Using the same asymmetric cutoff as for the thermodynamic potential, we get
\begin{eqnarray}
\frac{\cal P}{N} & = & \int_{-\Lambda-\mu_R}^{\Lambda+\mu_L} \frac{dk}{2\pi} \left( k - \mu \frac{k}{\epsilon_k} + \mu_5 \right)
\nonumber \\
& = & \int_{-\Lambda-\mu_R}^{\Lambda+\mu_L} \frac{dk}{2\pi} k + \mu \frac{\rho_5}{N} + \mu_5 \frac{\rho}{N}
\nonumber \\
& = & \frac{\mu \mu_5}{\pi}.
\label{A22}
\end{eqnarray}
Let us compute the invariant mass of a chirally imbalanced chunk of matter with size $L$,
\begin{eqnarray}
E & = & N L \frac{\mu^2+\mu_5^2}{2\pi}
\nonumber \\
P & = &  N L \frac{\mu \mu_5}{\pi}
\nonumber \\
M^2 & = &  E^2-P^2  = N^2 L^2 \left( \frac{ \mu^2-\mu_5^2}{2\pi} \right)^2
\label{A23}
\end{eqnarray}
The invariant mass $M$ is  indeed a Lorentz scalar, proportional to  $|j_{\mu}j^{\mu}|= |\rho^2-\rho_5^2|$.
These results coincide with what one would expect for a free gas of massless fermions with 
different densities of  left- and right handed fermions. Indeed, there one fills all positive energy levels using asymmetric Fermi surfaces (see Fig.~\ref{fig3})
\begin{eqnarray}
\rho_R & = & \int_0^{\mu_R} \frac{dk}{2\pi}  = \frac{\mu_R}{2\pi}= \frac{\mu+\mu_5}{2\pi},
\nonumber \\
\rho_L & = &  \int_{-\mu_L}^0 \frac{dk}{2\pi} = \frac{\mu_L}{2\pi}= \frac{\mu - \mu_5}{2\pi},
\nonumber \\
{\cal E} & = & \int_{-\mu_L}^{\mu_R} \frac{dk}{2\pi} |k| = \frac{\mu_R^2 + \mu_L^2}{4\pi} = \frac{\mu^2+\mu_5^2}{2 \pi},
\nonumber \\
{\cal P} & = & \int_{-\mu_L}^{\mu_R} \frac{dk}{2\pi} k = \frac{\mu_R^2-\mu_L^2}{4\pi} = \frac{\mu \mu_5}{\pi}.
\label{A24}
\end{eqnarray}
Vacuum subtraction is trivial here and amounts to ignoring negative energy states of the Dirac sea. The coincidence in all observables between
the free, massless Fermi gas and the chiral spiral is non-trivial, since the integration limits, the integrands and the necessary vacuum subtraction
are all different in the two cases.

\begin{figure}
\begin{center}
\epsfig{file=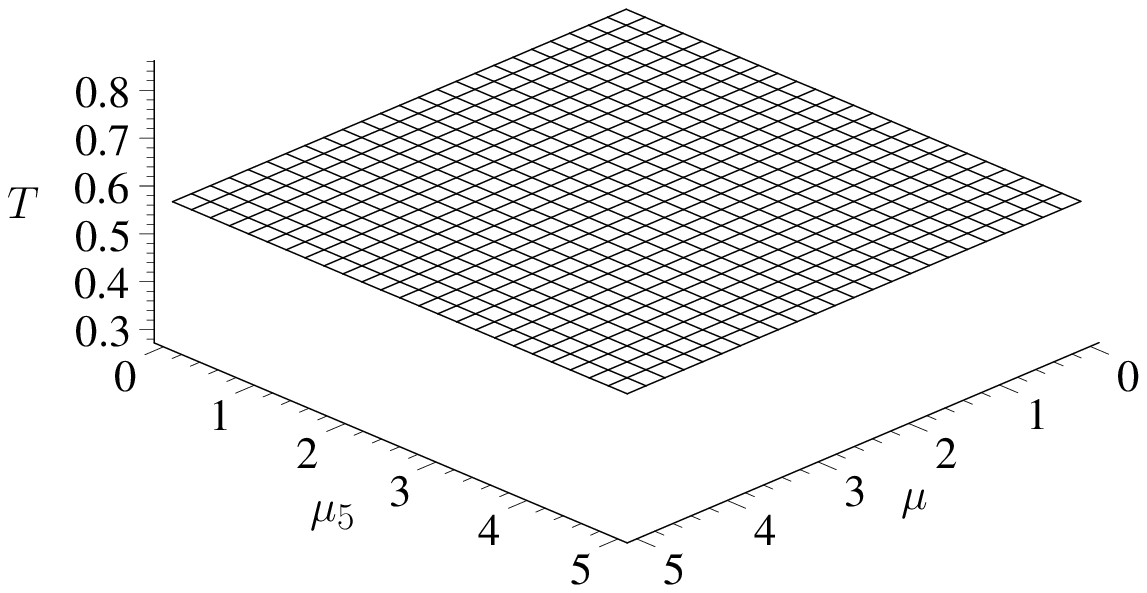,width=8cm,angle=0}
\caption{Full phase diagram of NJL$_2$ model in ($\mu, \mu_5, T$) space, in units where $m_0=1$. There is a single, horizontal critical sheet at $T=T_c$ where chiral symmetry gets restored.
Below this sheet, the mean field has the form of the chiral spiral, independently of $\mu_5$}
\label{fig4}
\end{center}
\end{figure}

In Fig.~\ref{fig4} we map out the phase diagram in ($\mu,\mu_5,T$) space. Like at $\mu_5=0$, the radius of the chiral spiral depends only on $T$, the pitch only on $\mu$. 
The critical temperature where chiral symmetry is restored in a 2nd order phase transition, $T_c=m_0e^{\gamma_E}/\pi$, is independent of both
chemical potentials. No phase transition occurs as a function of $\mu$ or $\mu_5$. The resulting phase diagram is therefore extremely simple, exhibiting
a single critical surface in the form of a horizontal plane (Fig.~\ref{fig4}). Above this sheet, chiral symmetry is restored and we are dealing with a hot Fermi gas 
of non-interacting, massless particles. Below the sheet, the order parameter has the chiral spiral form (\ref{A8}) with $m$ the temperature dependent mass, irrespective of  $\mu_5$.
The fermion densities depend on $\mu, \mu_5$, but not on temperature. If we move inside this region with the chiral spiral order parameter, there are nevertheless
observables which depend on all three variables ($\mu, \mu_5, T$). This will be discussed in greater detail in Sect.~\ref{sect4}.

Finally, let us briefly comment on the relationship between the present work and Ref.~\cite{9}. The authors of \cite{9} work out the phase diagram
of an extended NJL$_2$ model including a Cooper pairing interaction. They use as variational ansatz chiral spiral type potentials, both for the 
quark-antiquark and diquark condensates. In case the quark-antiquark pairing is stronger than diquark pairing, they find that the diquark condensate 
vanishes. But this means that there is no difference between the original NJL$_2$ model (\ref{I1}) and the extended model, at least at the mean field level.
Thus we can compare our results directly to those of Ref.~\cite{9}. We find that the value of the effective potential and the densities
$\rho, \rho_5$ agree perfectly. This is also true for the observation that the ($\mu,\mu_5$) and temperature dependences decouple in all observables. 
What we can add to Ref.~\cite{9} is the result that this is a fully self-consistent HF solution, rather than a variational approximation, and the physical 
picture as explained in Sect.~\ref{sect4}.
 
\section{Relation to time dependent approach at zero temperature}
\label{sect3}

Since a dense system carries both density and current density if viewed from a moving Lorentz frame, it is plausible that one can also deal with chiral
imbalance by applying a boost to a chirally symmetric system. This is the attitude taken in Ref.~\cite{10}. If one boosts the standard chiral spiral potential, it evidently becomes time dependent.
It is remarkable that the results of the time dependent approach in Ref.~\cite{10} and the present static one for energy density and momentum density agree
perfectly, yielding the correct relativistic energy-momentum relation for a finite piece of matter. This points to an ambiguity of the mean field. The origin of 
this ambiguity is the subject of the present section. 

In essence, the difference between Ref.~\cite{10} and  the present approach (at $T=0$) is the unitary transformation used to ``gauge away" the chemical potentials.
Suppose we start from the time dependent HF equation rather than from (\ref{A1})
\begin{equation}
\left( - \gamma_5 i \partial_x - \mu - \mu_5 \gamma_5 + \gamma^0 S+ i \gamma^1 P\right) \psi_{\alpha} = i \partial_t  \psi_{\alpha}.
\label{B1}
\end{equation}
One can eliminate $\mu_5$ either by the static transformation $\exp (i\mu_5 x)$ as in  Eq.~(\ref{A3}), or by a time dependent
axial transformation $\exp (i\mu_5 t \gamma_5)$. As far as $\mu$ is concerned, there would be a time dependent option as well, but we stick to the static choice
$\exp(i \mu x \gamma_5$) for the present purpose.
The results for fermion densities, energy and momentum do not seem to depend on the particular choice, but the order parameter (\ref{A8}) acquires a periodic time dependence
if one follows Ref.~\cite{10},
\begin{equation}
\Delta = S-iP = e^{2i(\mu x + \mu_5 t)}.
\label{B2}
\end{equation}
This expression can be interpreted as a boosted chiral spiral as follows. We have to distinguish the cases $|\mu_5| <> |\mu|$.
Starting from a chirally symmetric system with $\mu=\mu^{(0)}, \mu_5=0$ and boosting it, we find
\begin{equation}
\Delta  =  e^{2i\mu^{(0)} \gamma(x-vt)}  , \quad \gamma = (1-v^2)^{-1/2}.
\label{B3}
\end{equation}
Matching (\ref{B3}) to (\ref{B2}) yields
\begin{equation}
\mu^{(0)} = {\rm sgn}\,(\mu)\sqrt{\mu^2-\mu_5^2} , \quad v = - \frac{\mu_5}{\mu}.
\label{B4}
\end{equation}
Likewise, starting from a maximally imbalanced system with $\mu=0, \mu_5 = \mu_5^{(0)}$ and boosting it, we find
\begin{equation}
\Delta  =  e^{2i\mu_5^{(0)} \gamma(t-vx)}  
\label{B5}
\end{equation}
with the matching relations
\begin{equation}
\mu_5^{(0)} = {\rm sgn}\,(\mu_5) \sqrt{\mu_5^2-\mu^2} , \quad v = -  \frac{\mu}{\mu_5}.
\label{B6}
\end{equation}
We have to face the puzzling situation that two qualitatively different mean fields in chirally imbalanced systems yield identical global
observables. Which one is the correct one?

In fact, we believe that both approaches are legitimate and reflect a symmetry of the NJL$_2$ model whose impact has not
yet been fully appreciated. The NJL$_2$ Lagrangian in 1+1 dimensions shares a well known ``half-local" symmetry with the theory of free,
massless Dirac fermions. The Lagrangian (\ref{I1}) is not only invariant under global chiral transformations, 
\begin{equation}
\psi_R \to e^{i\alpha} \psi_R, \quad \psi_L \to e^{i \beta} \psi_L,
\label{B7}
\end{equation}
but also under a more general class of such transformations where $\alpha,\beta$ are arbitrary functions of one light cone variable each,
\begin{equation}
\psi_R \to e^{i \alpha(x-t)} \psi_R, \quad \psi_L  \to e^{i \beta(x+t)} \psi_L.
\label{B8}
\end{equation}
This symmetry is somewhat hidden in (\ref{I1}) but becomes manifest upon introducing
chiral spinor components and light cone coordinates
\begin{equation}
z  =  x-t,\ \ \bar{z}=x+t, \ \  \partial_0 = \bar{\partial}-\partial, \ \  \partial_1 = \bar{\partial} + \partial.
\label{B9}
\end{equation}
The Lagrangian (\ref{I1}) then assumes the form 
\begin{equation}
{\cal L} = 2i \psi_R^{\dagger} \bar{\partial} \psi_R - 2 i \psi_L^{\dagger} \partial \psi_L + 2 g^2 (\psi_L^{\dagger} \psi_R)(\psi_R^{\dagger}\psi_L)
\label{B10}
\end{equation}
and is clearly invariant under 
\begin{equation}
\psi_R \to e^{i \alpha(z)} \psi_R, \quad \psi_L \to e^{i \beta(\bar{z})} \psi_L.
\label{B11}
\end{equation}
In Euclidean space where $z,\bar{z}$ are complex conjugate, the corresponding symmetry transformation is referred to as ``left holomorphic" and ``right antiholomorphic" chiral 
transformations. It plays an important role in conformal field theory, affine current algebra, bosonization and is at the origin of the fact that $\rho_5=j$ in 1+1 dimensions \cite{13}.
It also has been found to be relevant in the context of two dimensional gauge theories on the light cone, where it has been invoked to explain the appearance of massless baryons
in the 't~Hooft model \cite{14}. 

Now consider the quotient of the unitary factors used in the present work ($U_I$) and in Ref.~\cite{10} ($U_{II}$),
\begin{eqnarray}
U_I  =  e^{i\mu x \gamma_5} e^{i \mu_5 x} & , & U_{II} = e^{i\mu x \gamma_5} e^{i \mu_5 t \gamma_5}
\nonumber \\
U_{II}^{\dagger} U_{I}  & = & e^{i \mu_5 (x - \gamma_5 t)}.
\label{B13}
\end{eqnarray}
This fits nicely into the form of the symmetry (\ref{B8}). The phase of the mean field $\Delta=S-iP$ is not invariant under transformation (\ref{B8}).  
It is well known that a constant phase is not observable, due to invariance of the theory under global chiral rotations (\ref{B7}). Now it seems that there is much
more freedom in the choice of the phase than that. This is somewhat disconcerting and may force us to reconsider more carefully the question to what 
extent the phase of the mean field $\Delta=S-iP$ is observable at all.

As a last remark, we come back to Ref.~\cite{9}. The authors emphasize the duality between quark-antiquark and quark-quark condensation, or chiral symmetry breaking and 
Cooper pairing, originally pointed out in \cite{15}. If we apply the appropriate canonical transformation (i.e., a particle-hole conjugation for left handed quarks only) to
the NJL$_2$ model, we can translate all our findings to a model featuring superconductivity rather than chiral symmetry breaking. Up to the interchange of $\mu$ and $\mu_5$
and a reinterpretation of the chiral condensate as Cooper pair condensate, all results carry over in a one-to-one fashion. This implies for instance the coexistence of static and
time dependent chiral spiral realizations of the Cooper pair model with two chemical potentials. 

From a practical point of view, the present static description has the advantage that it is easier to generalize to finite temperature, using only conventional
formalism. Hence we shall stick to the static formulation here.

\section{Physical picture of the chiral spiral}
\label{sect4}

Although the HF solution presented in Sect.~\ref{sect2} seems to be formally correct, it is not easy to interpret in terms of physics. Let us briefly come back to Figs.~\ref{fig1} and \ref{fig2}.
Curve 1 in either figure shows the spectrum and occupation of the negative energy vacuum levels, i.e., the filled Dirac sea. As is well known, the correct physical interpretation
requires to redefine occupied and unoccupied states for negative energy levels by a particle-hole conjugation. A hole in the filled Dirac sea is an antiparticle (``antiquark"), 
whereas particles (``quarks") correspond to occupied positive energy levels. Thus the vacuum contains neither quarks nor antiquarks, a precondition for its 
Lorentz invariance. The traditional HF picture of dense matter would suggest filling a number of positive energy levels in addition 
to the sea. Vacuum subtraction then simply amounts to ignore the fully occupied negative energy levels. The picture implied by the chiral spiral is radically different (Fig.~\ref{fig1}),
suggestive of adding occupied negative energy states in the UV region by extending the cutoff. How do we have to interpret this state? What is the correct definition
of quarks and antiquarks, and how do we subtract unobservable vacuum effects? At this point, this is still very unclear. The same holds true once we allow for chiral imbalance,
as in curve 3 of Fig.~\ref{fig2}.

To clarify the physics, we propose to compute observables which give more detailed information about the 
structure of dense matter, namely momentum distributions of quarks and antiquarks. Since this has not yet been done before, we start out with the 
chirally symmetric case ($\mu_5$=0) and indicate changes due to chiral imbalance later on.

To this end, we cast the transition (\ref{A3}) between chiral spiral and vacuum spinors into the form of a Bogoliubov transformation, following Ref. \cite{16}. 
Recall that the HF equation for the system with chemical potential $\mu$ is
\begin{equation}
\left( - \gamma_5 i \partial_x - \mu  + \gamma^0 S+ i \gamma^1 P\right) \psi_k^{(\pm)} = \pm \epsilon_k \psi_k^{(\pm)}
\label{C1}
\end{equation}
with 
\begin{equation}
S = m \cos 2\mu x, \quad P = -m \sin 2 \mu x, \quad \epsilon_k = \sqrt{k^2+m^2}.
\label{C2}
\end{equation}
The transformation
\begin{equation}
\psi_k^{(\pm)} = U \phi_k^{(\pm)}, \quad U = e^{i \mu x \gamma_5} 
\label{C3}
\end{equation}
maps this onto the vacuum HF equation
\begin{equation}
\left( - \gamma_5 i \partial_x  + \gamma^0 m \right) \phi_k^{(\pm)} = \pm \epsilon_k \phi_k^{(\pm)}.
\label{C4}
\end{equation}
The explicit form of the vacuum spinors in the representation
\begin{equation}
\gamma^0 = - \sigma_1, \quad \gamma^1 = i \sigma_3, \quad \gamma_5 = \gamma^0 \gamma^1 = - \sigma_2
\label{C5}
\end{equation}
is
\begin{equation}
\phi_k^{(+)}  =  u_k e^{ikx}, \quad \phi_k^{(-)} =  v_k e^{ikx}
\label{C6}
\end{equation}
where
\begin{equation}
u_k  =  \frac{1}{\sqrt{2} \epsilon_k} \left( \begin{array}{c} ik-m \\ \epsilon_k \end{array} \right), \ \ 
v_k = \frac{1}{\sqrt{2} \epsilon_k} \left( \begin{array}{c} ik-m \\ - \epsilon_k \end{array} \right).
\label{C7}
\end{equation}
The field operator $\Psi(x)$ (using the Schr\"odinger picture) can be expanded either using free spinors $\phi_k^{(\pm)}$, or using the full HF spinors 
$\psi_k^{(\pm)}$ 
\begin{eqnarray}
\Psi(x) & = & \sum_k \left[ a_k \phi_k^{(+)}(x) + b_k \phi_k^{(-)}(x) \right]
\nonumber \\
& = & \sum_k \left[ A_k \psi_k^{(+)}(x) + B_k \psi_k^{(-)}(x) \right].
\label{C8}
\end{eqnarray}
The relation between the 2nd quantized fermion operators $a_k,b_k$ and $A_k,B_k$ is then given by the Bogoliubov transformation
\begin{eqnarray}
a_k & = & \sum_{\ell} \left[ (\phi_k^{(+)},\psi_{\ell}^{(+)})A_{\ell} +  (\phi_k^{(+)},\psi_{\ell}^{(-)})B_{\ell} \right],
\nonumber \\
b_k & = & \sum_{\ell} \left[ (\phi_k^{(-)},\psi_{\ell}^{(+)})A_{\ell} +  (\phi_k^{(-)},\psi_{\ell}^{(-)})B_{\ell} \right].
\label{C9}
\end{eqnarray}
The quark momentum distribution reads \cite{16}
\begin{equation}
W_q(k) = \langle {\rm HF}| a_k^{\dagger} a_k |{\rm HF} \rangle = \sum_{\ell}  | ( \phi_k^{(+)}, \psi_{\ell}^{(-)})|^2 .
\label{C10}
\end{equation}
Here we have used the fact that in the HF state, all $\psi^{(-)}$ states are occupied. 
For the antiquarks, we perform the particle-hole conjugation as usual and get
\begin{equation}
W_{\bar{q}}(k) = 1- \langle {\rm HF}| b_{-k}^{\dagger} b_{-k} |{\rm HF} \rangle =1- \sum_{\ell}  | ( \phi_{-k}^{(-)}, \psi_{\ell}^{(-)})|^2 .
\label{C11}
\end{equation}
The relevant Bogoliubov coefficients can easily be evaluated. 
We decompose $U$ with the help of R/L projection operators
\begin{equation}
U = e^{i \mu x \gamma_5} = P_R e^{i \mu x} + P_L e^{-i \mu x}, \quad P_{R,L} = \frac{1 \pm \gamma_5}{2}
\label{C12}
\end{equation}
and find
\begin{eqnarray}
(\phi_k^{(+)}, \psi_{\ell}^{(-)})  
& = & \delta_{\ell,k-\mu} u_k^{\dagger} P_R v_{\ell} + \delta_{\ell,k+ \mu} u_k^{\dagger} P_L v_{\ell},
\nonumber  \\
(\phi_{-k}^{(-)}, \psi_{\ell}^{(-)}) 
& = & \delta_{\ell,-k- \mu} v_{-k}^{\dagger} P_R v_{\ell} 
\nonumber \\
& & 
+ \delta_{\ell,-k+\mu} v_{-k}^{\dagger} P_L v_{\ell}.
\label{C13} 
\end{eqnarray}
Consequently
\begin{eqnarray}
W_q(k)  & = & |u_k^{\dagger}P_R v_{k-\mu}|^2 +|u_k^{\dagger} P_L v_{k+\mu}|^2,
\nonumber  \\
W_{\bar{q}}(k)  & = & 1 - |v_{-k}^{\dagger}P_R v_{-k-\mu}|^2 
\nonumber \\
& & 
-|v_{-k}^{\dagger} P_L v_{-k+\mu}|^2.
\label{C14}
\end{eqnarray}
These expressions are only valid at $\mu \neq 0$. If $\mu=0$, the 
two terms on the right hand side of (\ref{C13}) have to be added coherently before squaring, so that 
both distribution functions vanish identically. For $\mu \neq 0$, a
simple computation using the explicit spinors (\ref{C7}) yields
\begin{eqnarray}
W_q(k) & = &  \frac{1}{2} - \frac{V_{k-\mu}(1+V_k)}{4} + \frac{V_{k+\mu}(1-V_k)}{4},
\nonumber \\
W_{\bar{q}}(k) & = & \frac{1}{2} - \frac{V_{k+\mu}(1+V_k)}{4} + \frac{V_{k-\mu}(1-V_k)}{4},
\label{C15}
\end{eqnarray}
where
\begin{equation}
V_p = \frac{p}{\epsilon_p} = \frac{\partial \epsilon_p}{\partial p}
\label{C16}
\end{equation}
denotes the velocity of a free, massive fermion with momentum $p$.
Notice that a change of sign of $\mu$ interchanges $W_q$ and $W_{\bar{q}}$, as
expected. Both distribution functions are even under $k\to -k$.
Actually, we can further decompose $W_q, W_{\bar{q}}$ according to chirality. The factors
$(1 \pm V_k)/2$ occuring in (\ref{C15}) can be used to ``tag" the $L/R$ components, since
\begin{equation}
u_k^{\dagger} \gamma_5 u_k  =   v_{-k}^{\dagger} \gamma_5  v_{-k} = V_k
\label{C18}
\end{equation} 
All we have to do to disentangle $R/L$ contributions to $W_q, W_{\bar{q}}$ is to decompose the constant 1/2 into
\begin{equation}
\frac{1}{2} = \frac{1}{2} \left( \frac{1+V_k}{2} + \frac{1-V_k}{2} \right) 
\label{C19}
\end{equation}
The result of this decomposition are momentum distributions for $R/L$ quarks/antiquarks,
\begin{eqnarray}
W_q^R(k) & = & \left( \frac{1+V_k}{2} \right) \left( \frac{1-V_{k-\mu}}{2}\right),
\nonumber \\
 W_q^L(k) & = &  \left( \frac{1-V_k}{2} \right) \left( \frac{1+V_{k+\mu}}{2} \right),
\nonumber \\
W_{\bar{q}}^R(k) & = & \left( \frac{1+V_k}{2} \right) \left( \frac{1-V_{k+\mu}}{2}\right),
\nonumber \\
W_{\bar{q}}^L(k) & = &  \left( \frac{1-V_k}{2} \right) \left( \frac{1+V_{k-\mu}}{2}\right).
\label{C20}
\end{eqnarray}
Changing the sign of $\mu$ interchanges $q/\bar{q}$, changing the sign of $k$ interchanges $R/L$.

\begin{figure}
\begin{center}
\epsfig{file=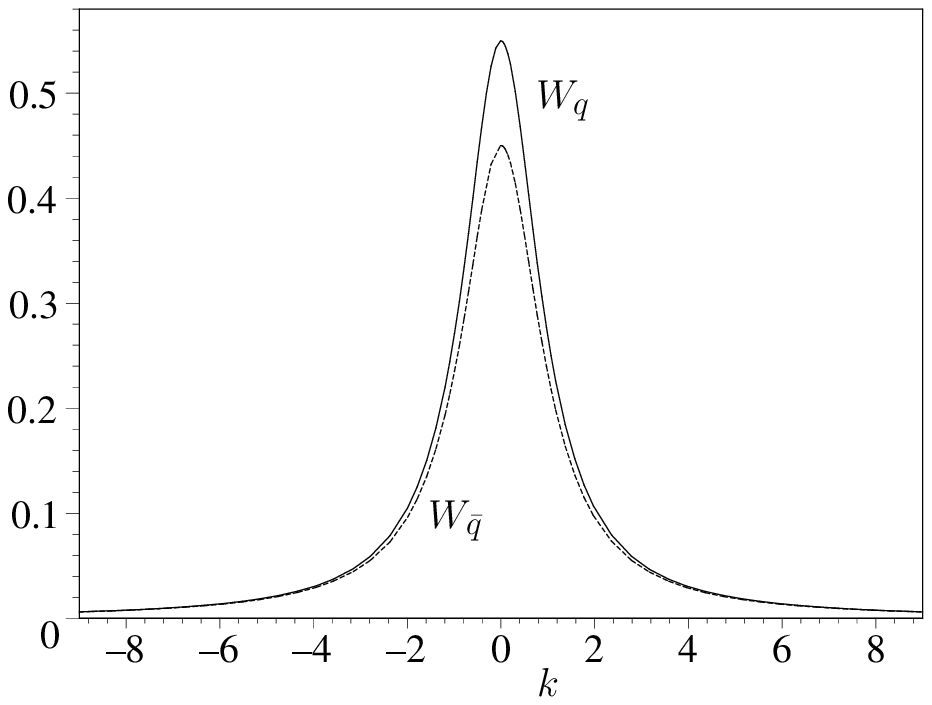,width=8cm,angle=0}
\caption{Momentum distributions of quarks ($W_q$) and antiquarks ($W_{\bar{q}}$), Eq.~(\ref{C15}), for a system with low density ($m=1,\mu=0.1,\mu_5=0$).}
\label{fig5}
\end{center}
\end{figure}

\begin{figure}
\begin{center}
\epsfig{file=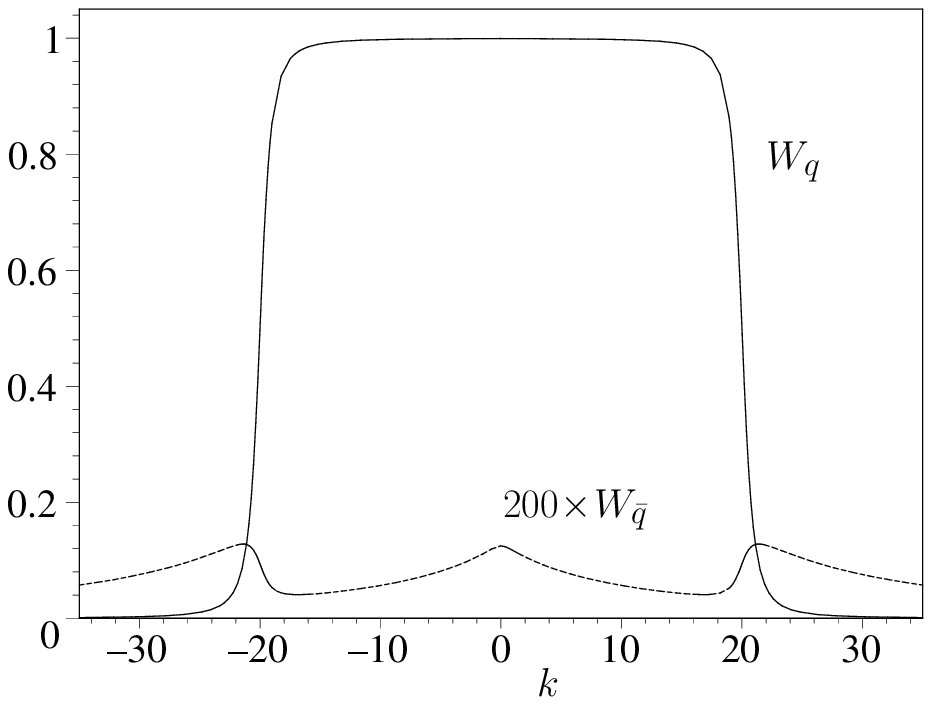,width=8cm,angle=0}
\caption{Same as Fig.~5, but at high density ($m=1, \mu=20, \mu_5=0$).}
\label{fig6}
\end{center}
\end{figure}

We first illustrate the distribution functions at low (Fig.~\ref{fig5}) and high (Fig.~\ref{fig6}) densities.
Both figures show immediately that the structure of chiral spiral matter has nothing to do with 
the UV region. At the smaller density, we see a strongly relativistic signature with almost equal densities 
of rather low momentum quarks and antiquarks. Actually, the limit $\mu \to 0$ is
\begin{equation}
\lim_{\mu \to 0} W_q= \lim_{\mu \to 0} W_{\bar{q}} = W_0 =  \frac{m^2}{2(m^2+k^2)} .
\label{C23}
\end{equation}
This does not describe the vacuum where all distribution functions vanish, but should be thought of as the momentum distribution in the low density limit, i.e., for a single,
massless, delocalized baryon. As pointed out in Refs.~\cite{17} and \cite{5}, the baryon in the NJL$_2$ picture is the simplest realization of the 
Skyrme model \cite{18} where baryon number arises from winding number of the pion field. This is supported by the equal distributions of quarks and 
antiquarks as well as by the momentum distribution (\ref{C23}) which agrees with the absolute square of the pion Bethe--Salpeter amplitudes \cite{3} up to a 
normalization factor. At high densities (Fig.~\ref{fig6}), the picture we get looks surprisingly familiar. To a good approximation, we can neglect antiquarks
and recover the picture where quark levels are filled up to a Fermi momentum. The only difference is the fact that the Fermi surface is not sharp
but rounded off, similar to what happens at finite temperature. This picture is quite different from the impression conveyed superficially by curve 2 in Fig.~\ref{fig1}.

\begin{figure}
\begin{center}
\epsfig{file=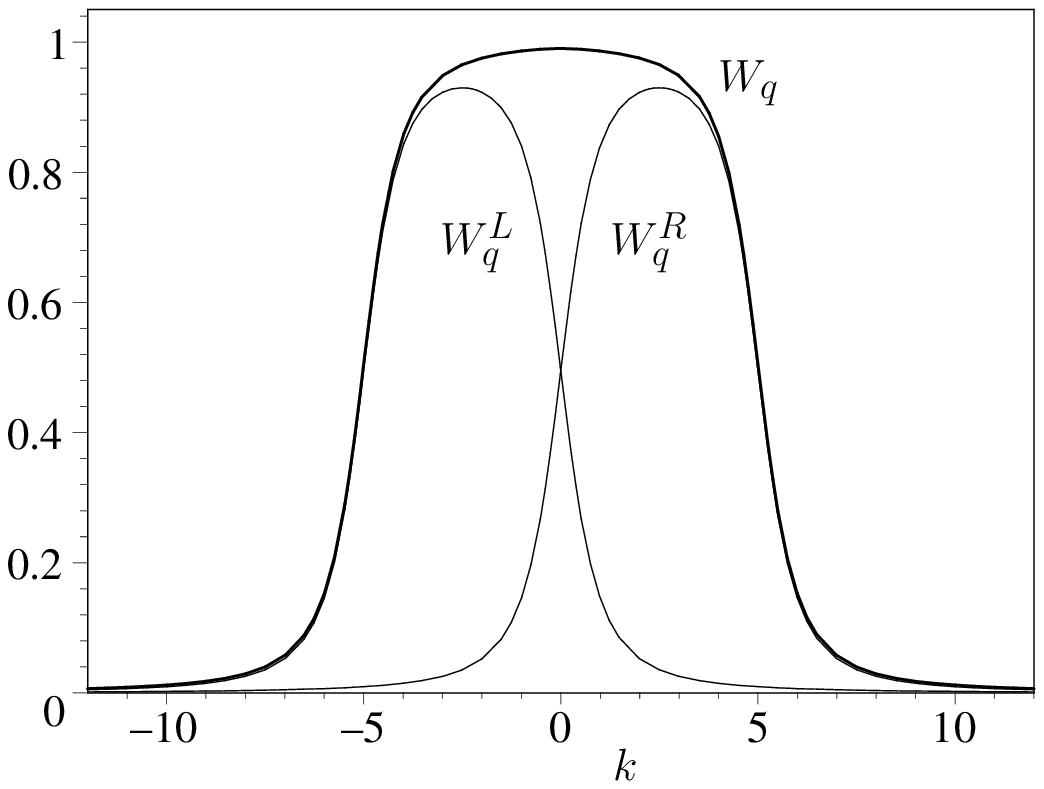,width=8cm,angle=0}
\caption{Splitting up $W_q$ into contributions with different chiralities $R/L$, see Eq.~(\ref{C20}). A case of intermediate density ($m=1, \mu=5, \mu_5=0$) is shown.}
\label{fig7}
\end{center}
\end{figure}

\begin{figure}
\begin{center}
\epsfig{file=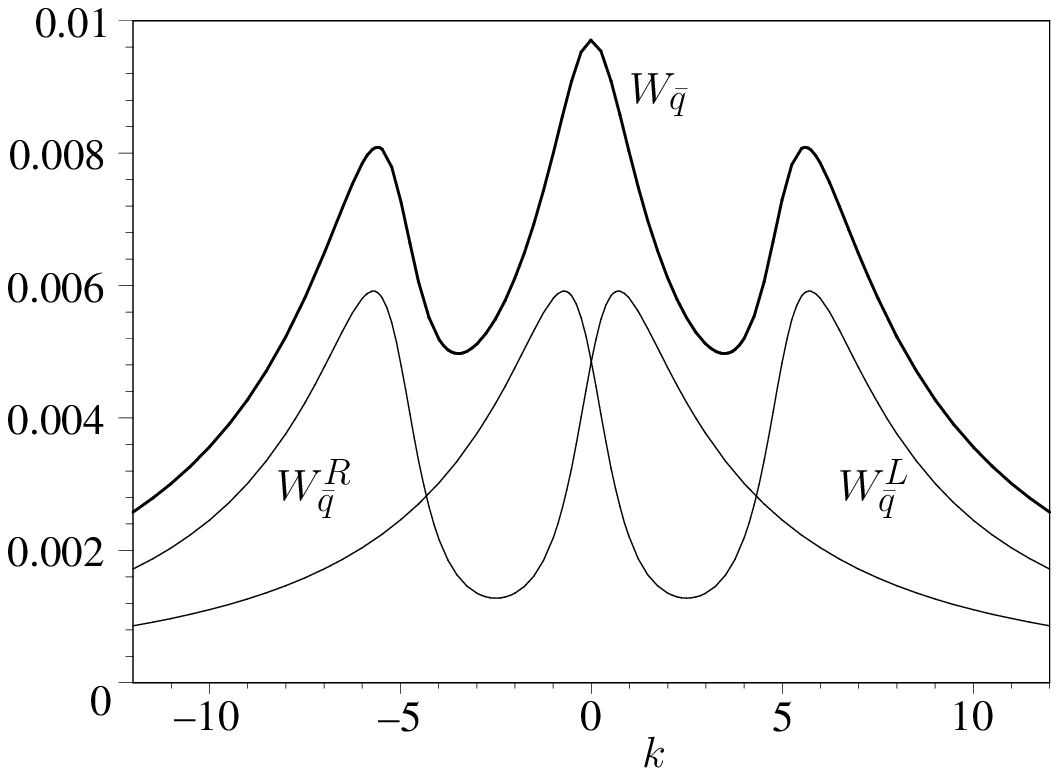,width=8cm,angle=0}
\caption{Same as Fig.~\ref{fig7}, but for antiquark distribution $W_{\bar{q}}$.}
\label{fig8}
\end{center}
\end{figure}

Figs.~\ref{fig7} and \ref{fig8} show how the quark and antiquark momentum distributions split up according to chirality into $R/L$ pieces. We have plotted the
distributions from Eq.~(\ref{C20}) for a density in between the two cases shown in Figs.~\ref{fig5}, \ref{fig6}. Notice that quarks and antiquarks with the same chirality 
move predominantly into opposite directions. 

An important cross check of the whole formalism is the evaluation of global observables in this new setting. The fermion density 
can be evaluated analytically in a straightforward way, without introducing any UV cutoff,
\begin{equation}
\frac{\rho}{N} = \int \frac{dk}{2\pi} \left( W_q-W_{\bar{q}} \right) = \frac{\mu}{\pi}.
\label{C24}
\end{equation}
This agrees with the result in Sect.~\ref{sect2}, confirming that the cutoff chosen there was reasonable. In the present calculation, the vacuum subtraction
happens when we perform the ph-conjugation for negative energy levels, Eq.~(\ref{C14}).
In order to compute the energy density, it is sufficient to subtract at $\mu=0$ to get a finite result (the $\mu=0$ part is eveluated differently and not of interest here.)
Thus
\begin{eqnarray}
\frac{{\cal E}(\mu) - {\cal E}(0)}{N} & = &  \int \frac{dk}{2\pi} \epsilon_k  \left(W_q+W_{\bar{q}}- 2 W_0 \right)
\nonumber \\
& = &  \frac{\mu^2}{2 \pi}.
\label{C26}
\end{eqnarray}
Once again, no cutoff is needed and we confirm our previous result analytically. 

\begin{figure}
\begin{center}
\epsfig{file=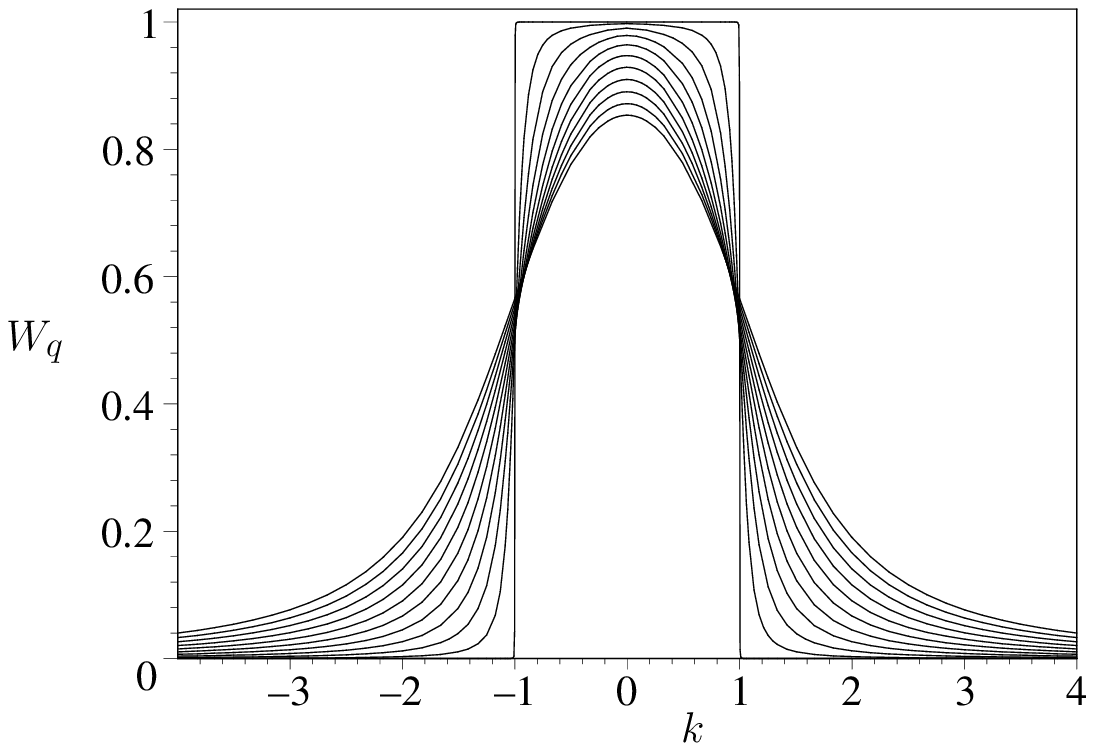,width=8cm,angle=0}
\caption{Evolution of quark momentum distribution with dynamical mass $m$, to illustrate restoration of chiral symmetry as a function of temperature.
Curves shown correspond to $\mu=1, \mu_5=0$ and varying $m$ in steps of 0.1 between 1 (smoothest curve) and 0 (rectangular curve).}
\label{fig9}
\end{center}
\end{figure}

\begin{figure}
\begin{center}
\epsfig{file=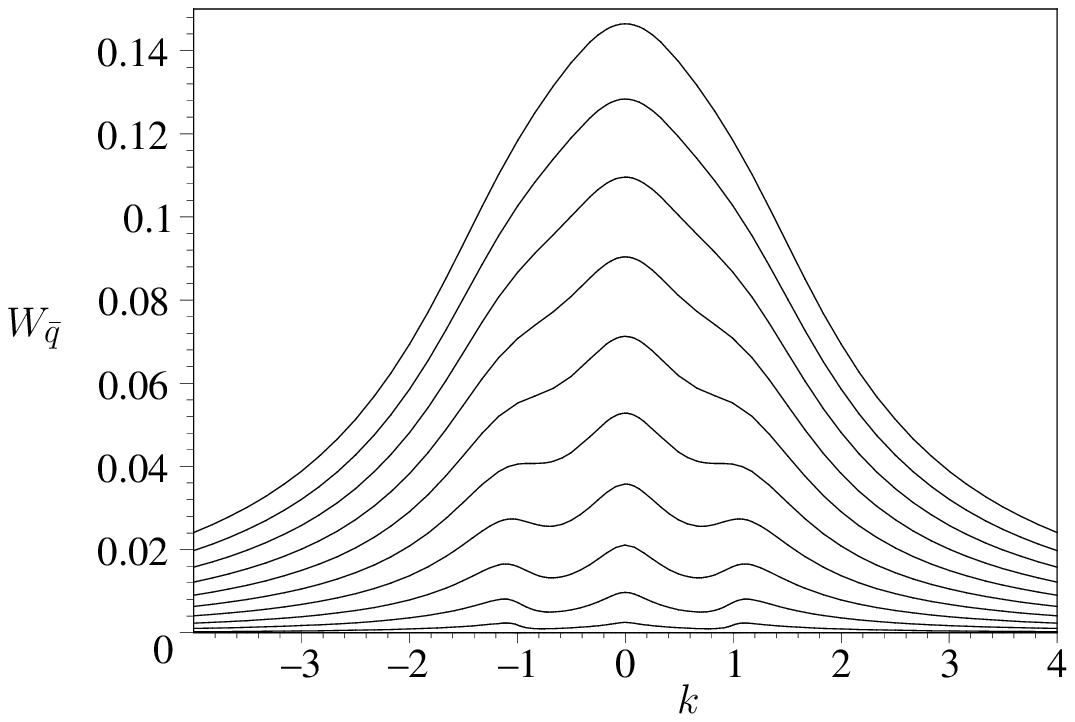,width=8cm,angle=0}
\caption{Like Fig.~\ref{fig9}, but for antiquarks. The height of the curves decreases as $m$ is varied from 1 to 0. Note the different scale as compared to Fig.~\ref{fig7}.}
\label{fig10}
\end{center}
\end{figure}

As a last illustration of the chirally symmetric results ($\mu_5=0$), we go back to the $\mu_5=0$ plane of the phase diagram shown in Fig.~\ref{fig4}. 
If we move vertically inside this diagram (i.e., vary $T$ at constant $\mu$), the dynamical mass $m$ changes, vanishing at $T_c$. Figs.~\ref{fig9} and \ref{fig10}
show the evolution of quark and antiquark distribution functions at $\mu=1$ as $m$ decreases from $m=1$ at $T=0$ to $m=0$ at $T=T_c$. 
At $m=0$, $W_q$ goes over into the result for free, massless fermions ($\mu>0$)
\begin{equation}
W_q(k) = \Theta(\mu-k)- \Theta(-\mu-k), \quad W_{\bar{q}}(k) = 0 .
\label{C26a}
\end{equation}
Thus restoration of chiral symmetry is also reflected in the distribution functions. If we move along a horizontal path through the phase diagram (by increasing $\mu$ at fixed $T$),
the distribution also approaches a rectangular shape of increasing width. If we would plot it as a function of $k/\mu$, the result would look just like Figs.~\ref{fig9} and \ref{fig10}. 
The only difference is the fact that the limit of a free, massless Fermi gas, signalling the restoration of chiral symmetry, would be reached only at $\mu \to \infty$, as compared to a finite 
temperature $T_c$ before. 

\begin{figure}
\begin{center}
\epsfig{file=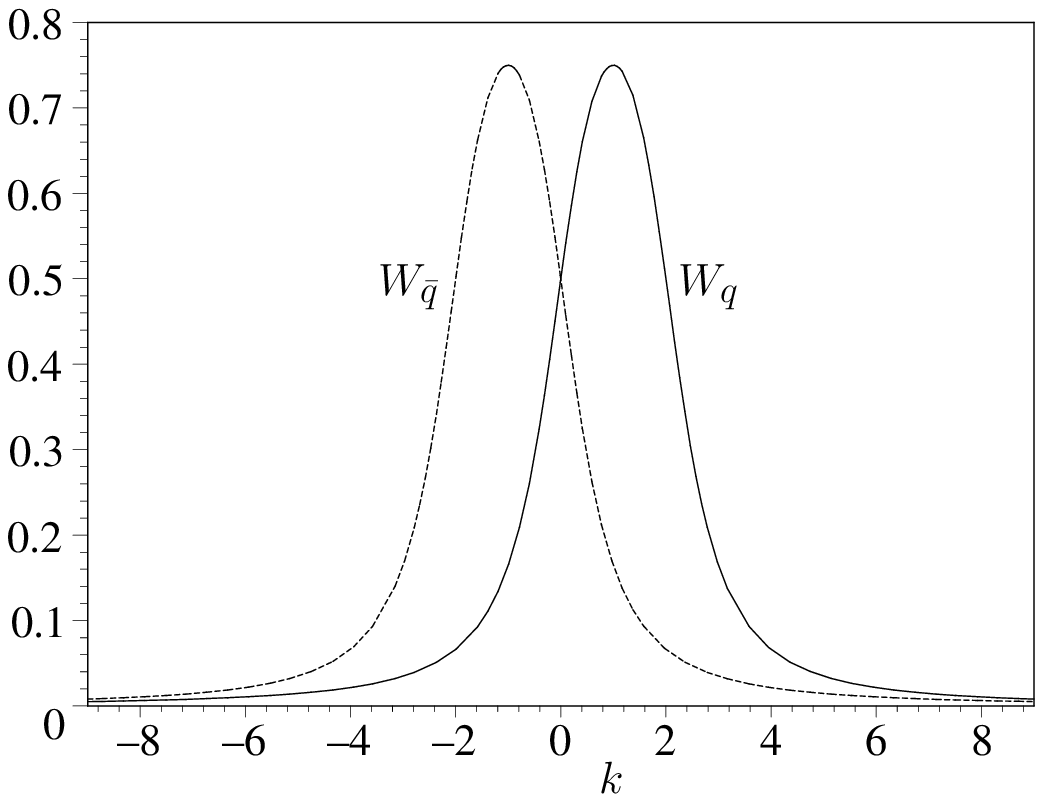,width=8cm,angle=0}
\caption{Quark and antiquark momentum distribution functions for a system with maximal chiral imbalance, Eq.~(\ref{C32}),
and rather low current density ($m=1,\mu=0, \mu_5=2$).}
\label{fig11}
\end{center}
\end{figure}

\begin{figure}
\begin{center}
\epsfig{file=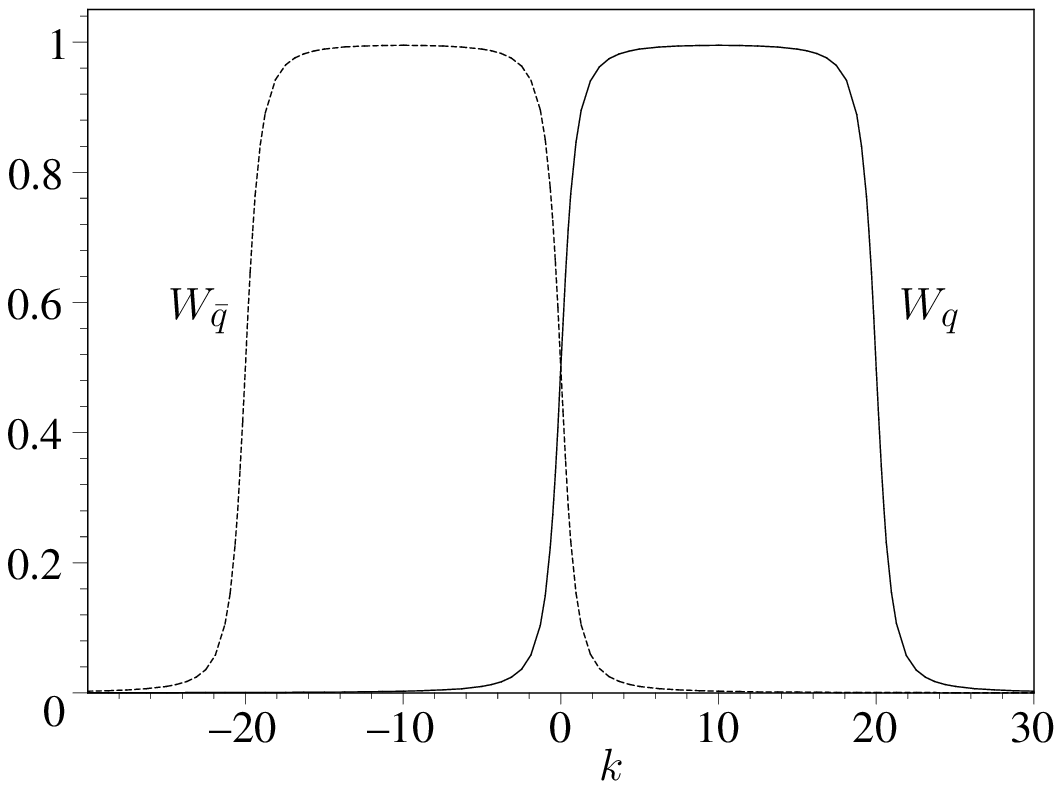,width=8cm,angle=0}
\caption{Like Fig.~\ref{fig11}, but for large current density ($m=1,\mu=0, \mu_5=20$).}
\label{fig12}
\end{center}
\end{figure}

What changes if we introduce $\mu_5$, allowing for chiral imbalance? Eq. (\ref{C1}) gets replaced by
\begin{equation}
\left( - \gamma_5 i \partial_x - \mu - \gamma_5 \mu_5 + \gamma^0 S+ i \gamma^1 P\right) \psi_k^{(\pm)} = \pm \epsilon_k \psi_k^{(\pm)}
\label{C27}
\end{equation}
and the unitary matrix $U$ in (\ref{C3}) by 
\begin{equation}
U = e^{i \mu x \gamma_5} e^{i \mu_5 x} = P_R e^{i \mu_R x} + P_L e^{-i \mu_L x}
\label{C28}
\end{equation}
Apart from the fact that the momentum splitting becomes asymmetric, the calculation of the Bogoliubov coefficients is the same as before.
The resulting quark and antiquark momentum distributions are
\begin{eqnarray}
W_q(k)  & = & |u_k^{\dagger}P_R v_{k-\mu_R}|^2 +|u_k^{\dagger} P_L v_{k+\mu_L}|^2,
\nonumber  \\
W_{\bar{q}}(k)  & = & 1 - |v_{-k}^{\dagger}P_R v_{-k-\mu_R}|^2 
\nonumber \\
& & 
-|v_{-k}^{\dagger} P_L v_{-k+\mu_L}|^2.
\label{C29}
\end{eqnarray}
or, explicitly,
\begin{eqnarray}
W_q(k) & = &  \frac{1}{2} - \frac{V_{k-\mu_R}(1+V_k)}{4} + \frac{V_{k+\mu_L}(1-V_k)}{4},
\nonumber \\
W_{\bar{q}}(k) & = & \frac{1}{2} - \frac{V_{k+\mu_R}(1+V_k)}{4} + \frac{V_{k-\mu_L}(1-V_k)}{4}.
\label{C30}
\end{eqnarray}
We can again decompose $W_q,W_{\bar{q}}$ into $R/L$ contributions
\begin{eqnarray}
W_q^R(k) & = & \left( \frac{1+V_k}{2} \right) \left( \frac{1-V_{k-\mu_R}}{2}\right), 
\nonumber \\
W_q^L(k) & = &  \left( \frac{1-V_k}{2} \right) \left( \frac{1+V_{k+\mu_L}}{2}\right),
\nonumber \\
W_{\bar{q}}^R(k) & = & \left( \frac{1+V_k}{2} \right) \left( \frac{1-V_{k+\mu_R}}{2}\right),
\nonumber \\
W_{\bar{q}}^L(k) &  = & \left( \frac{1-V_k}{2} \right) \left( \frac{1+V_{k-\mu_L}}{2}\right).
\label{C31}
\end{eqnarray}
Of particular interest is the simple special case of vanishing $\mu$ where $\mu_R = - \mu_L = \mu_5$, complementary
to the case of vanishing $\mu_5$ considered above. In this case, the HF potential is homogeneous ($S=m,P=0$), just as in 
the vacuum. Then
\begin{eqnarray}
W_q(k) & = & \frac{1}{2} \left( 1- V_k V_{k-\mu_5} \right)
\nonumber \\
W_{\bar{q}}(k) & = & \frac{1}{2} \left( 1- V_k V_{k+\mu_5} \right) = W_q(-k)
\label{C32}
\end{eqnarray}
The last symmetry follows from $V_p = - V_{-p}$, implies equal densities of quarks and antiquarks and is consistent with the 
fact that the momentum density vanishes if $\mu=0$, see Eq.~(\ref{A22}). 
The decomposition according to chirality can be inferred from (\ref{C31}) by specializing it to $\mu_R=-\mu_L=\mu_5$.
Examples of these distributions for small and large $\mu_5$ are shown in Figs.~\ref{fig11} and \ref{fig12}.
They actually look as expected. In order to produce a current at zero density, one combines a bunch of right moving
quarks with a bunch of left moving antiquarks. For $\mu_5 \to \infty$ or $m \to 0$, the distributions approach the expected rectangular 
ones for a free, massless Fermi gas. 

For the sake of completeness, we list the limit $m \to 0$, reached when approaching the critical
surface $T=T_c$ from below, for all distribution functions. Denoting the characteristic function of the interval $[a,b]$ on the $k$-axis
by
\begin{equation}
h(k,[a,b]) = \Theta(a-k)- \Theta(b-k) \quad (b>a),
\label{35}
\end{equation} 
we find
\begin{eqnarray}
W_q^R(k) & = & h(k,[0,\mu_R]), \ \ W_{\bar{q}}^R(k) = 0 \quad {\rm for\ }\mu_R>0,
\nonumber \\
W_{\bar{q}}^R(k) & = & h(k,[0,-\mu_R]), \ \  W_q^R(k) = 0 \quad {\rm for\ } \mu_R<0,
\nonumber \\
W_q^L(k) & = & h(k,[-\mu_L,0]), \ \  W_{\bar{q}}^L(k) = 0 \quad {\rm for\ } \mu_L>0,
\nonumber \\
W_{\bar{q}}^L(k) & = & h(k,[\mu_L,0]), \ \  W_q^L(k) = 0 \quad {\rm for\ } \mu_L>0.
\label{36}
\end{eqnarray}
All of these results agree with a free, massless Fermi gas with chiral imbalance.

\begin{figure}
\begin{center}
\epsfig{file=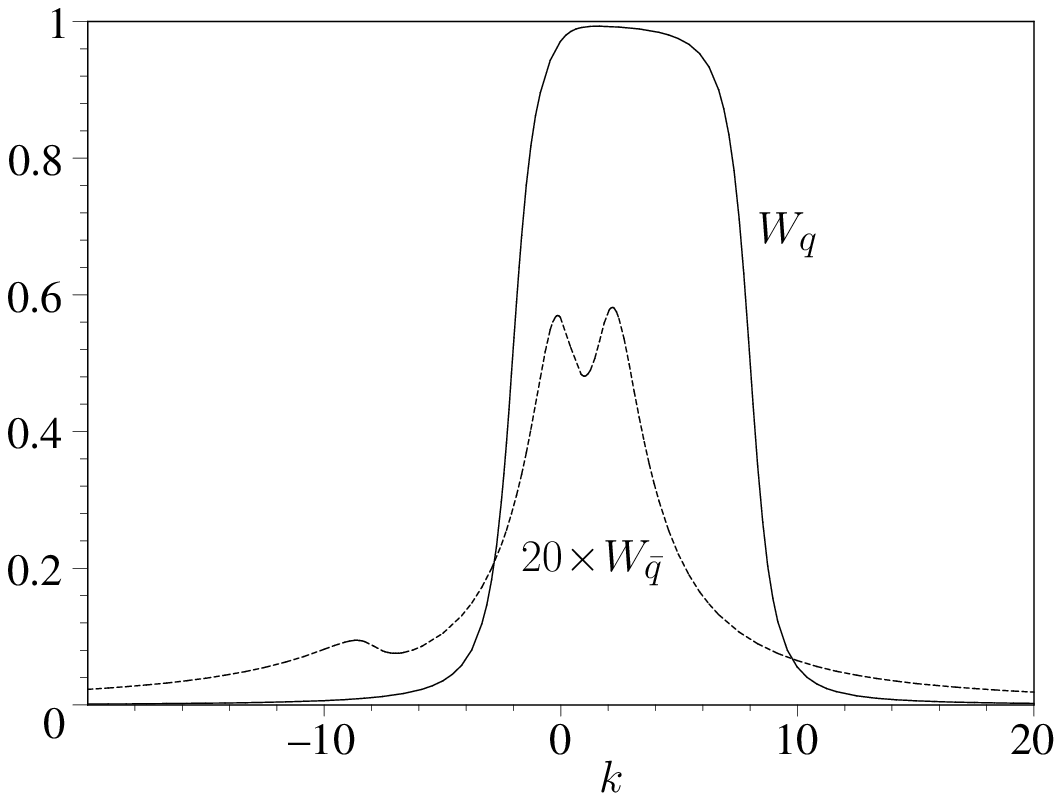,width=8cm,angle=0}
\caption{Example of quark and antiquark momentum distribution for the general case of  non-zero $\mu$ and $\mu_5$ ($m=1,\mu=5,\mu_5=3$), Eq.~(\ref{C30}).}
\label{fig13}
\end{center}
\end{figure}

If one allows both $\mu$ and $\mu_5$ to be non-zero, one gets distortions of these simple pictures. An example is shown in Fig.~\ref{fig13}. Other cases can easily be
generated with the help of the above analytical formulas.

Finally, we check the computation of observables in the case of chiral imbalance, a further test of the asymmetric cutoff introduced in Sect.~\ref{sect2}.
All calculations can be done in closed analytical form and do not require any cutoff, provided we subtract again the energy density at $\mu=\mu_5=0$.
We find 
\begin{eqnarray}
\frac{\rho_R}{N} & = & \int \frac{dk}{2\pi}\left( W_q^R- W_{\bar{q}}^R \right) = \frac{\mu_R}{2\pi},
\nonumber \\
\frac{\rho_L}{N} & = & \int \frac{dk}{2\pi}\left( W_q^L- W_{\bar{q}}^L \right) = \frac{\mu_L}{2\pi},
\nonumber \\
\frac{{\cal E}_R}{N} & = & \int \frac{dk}{2\pi}\epsilon_k \left( W_q^R + W_{\bar{q}}^R-  W_0 \right) = \frac{\mu_R^2}{4\pi},
\nonumber \\
\frac{{\cal E}_L}{N} & = & \int \frac{dk}{2\pi}\epsilon_k \left( W_q^L + W_{\bar{q}}^L-  W_0 \right) = \frac{\mu_L^2}{4\pi},
\nonumber \\
\frac{{\cal P}_R}{N} & = & \int \frac{dk}{2\pi} k \left( W_q^R + W_{\bar{q}}^R \right) = \frac{\mu_R^2}{4\pi},
\nonumber \\
\frac{{\cal P}_L}{N} & = & \int \frac{dk}{2\pi} k \left( W_q^L + W_{\bar{q}}^L \right) = - \frac{\mu_L^2}{4\pi},
\label{C34}
\end{eqnarray}
where all observables have been split into contributions from R/L fermions.
This fully confirms the simpler, but physically less transparent, cutoff calculation of Sect.~\ref{sect2}.

\section{Summary and conclusions}
\label{sect5}

In this work, we have presented the full phase diagram of the NJL$_2$ model in ($\mu, \mu_5, T$) space. Without chiral imbalance ($\mu_5=0$), it has been known 
for some time that the physics is strongly dominated by a chiral spiral type mean field. The radius of this helix structure in ($S,P,x$) space is determined by
the thermal mass, vanishing at a critical temperature in a continuous fashion. The period is determined by $\mu$. By generalizing this construction to 
finite $\mu_5$, we find that the mean field is unchanged, but that one has to choose an asymmetric cutoff in momentum space when summing over negative 
energy states. The resulting observables look very reasonable and are consistent with Lorentz covariance. To corroborate our choice of cutoff, we then
have evaluated for the first time momentum distribution functions for ``quarks" and ``antiquarks", using a standard Bogoliubov transformation.
On the one hand, this confirms the calculation of all global observables, now without need to introduce an UV cutoff. On the other hand, it sheds light
onto the structure of matter which is somewhat obscure in the original derivation. We find that at low densities, the distribution
functions reflect the close relation between baryon density and the pion field characteristic for the Skyrme picture. At high densities, one smoothly reaches the limit of a free, massless
Fermi gas. In the opposite case of $\mu=0, \mu_5 \neq 0$, one recovers the naively expected picture of equal numbers of right moving quarks and left moving
antiquarks, but for a homogeneous mean field. 

Perhaps the most surprising  finding is the fact that one can get the same global observables in this static way as in a time dependent approach, where the 
chiral spiral is boosted and develops a space-time dependence. The fact that both calculations agree on the global observables has been traced back to a half-local symmetry, well known
from massless free fermions but disregarded so far in the NJL$_2$ model. We do not believe that this ambiguity in the phase of the mean field $\Delta=S-iP$
renders our results for observables less reliable. However, one may have to rethink more fundamentally about how to deal with this half-local symmetry in a non-gauge theory,
and which quantities are really observable.

\end{document}